\documentstyle[aps,prb,floats]{revtex}
\begin{document}




{\bf Golubev and Zaikin reply:} In a Comment~\cite{AAG1} Aleiner,
Altshuler and Gershenzon (AAG) stated that
we~\cite{GZ1} ``found that ``zero-point fluctuations of electrons'' contribute
to the dephasing rate $1/\tau_{\varphi}$''. We made no such claims.
Obviously, the eigenmodes of any quantum system preserve their coherence.
On the other hand, single-particle properties of an interacting
system -- which are not directly expressed in terms of eigenmodes -- can
display a finite decoherence length $L_{\varphi}$ in the
low-temperature limit.
An example is provided by a particle coupled to a harmonic oscillator
bath~\cite{CL}, which we discuss explicitly below.
This is also the case for the weak localization correction to the conductivity
in the presence of interaction, where we found~ \cite{GZ1,GZ2}
$L_{\varphi}$ to remain finite at $T \to 0$.
AAG furthermore concluded that our procedure is ``simply wrong'' because ``some
contributions were lost'' during ``uncontrollable'' semiclassical averaging.
This is not the case, as demonstrated by the fact that
our approach \cite{GZ2} fully reproduces the results of
AAG \cite{AAG2} if analyzed on a perturbative level.
The advantage of our formulation is that we can proceed beyond.

The conductivity and the electron density matrix $\rho$
can be expressed in terms of a path integral \cite{GZ2}
\begin{equation}
\int {\cal D}\bbox{r}\int {\cal D}\bbox{p} \, \exp (iS_0-iS_0'-iS_R-S_I),
\label{1}
\end{equation}
where $S_0$ and $S_0'$ represent the electron action on the two parts of the
Keldysh contour, while $iS_R+S_I$ accounts for the interaction.
We reproduce AAG's results
on a perturbative level if we expand the path integral (\ref{1})
up to first order in $iS_R+S_I$.
To show the equivalence we perform (see Appendix A for details)
some exact transformations starting
from the formula for the conductivity of AAG \cite{AAG2},
making use of the expression for the
Green-Keldysh function
$
G^K_{\bbox{r}_1,\bbox{r}_2}(\epsilon)=\int d\bbox{r}'
\big[G^R_{\bbox{r}_1,\bbox{r}'}(\epsilon)-
G^A_{\bbox{r}_1,\bbox{r}'}(\epsilon)\big]
\big[\hat 1-2\rho_{\bbox{r}',\bbox{r}_2}
\big].
$
The result (of either approach) consists of the two groups of terms:
$G^RG^AG^AG^A$ and $G^RG^AG^A(1-2\rho )G^A$. Terms of the type
$G^RG^AG^RG^A$ (claimed to be lost~\cite{AAG2} in our calculation)
vanish due to the causality principle.
Thus no diagrams or paths are missing in our analysis.

The crucial difference between AAG's \cite{AAG2} and our \cite{GZ1,GZ2}
procedures is that we {\it do not} expand in $iS_R+S_I$,
but rather evaluate the full path integral (\ref{1}).
The expressions $S_R$ and $S_I$ are non-local in time and are not small at low
$T$ even if the interaction is weak. Hence an expansion of
the path integral (\ref{1}), which is equivalent to a
Golden-rule-type perturbation theory in lowest order in the
scattering processes,  becomes insufficient at low $T$.
The only approximation we employ is an RPA expansion of the
effective action. This yields tractable expressions for $S_R$ and
$S_I$, which include (within RPA) processes of all orders.
Averaging over disorder plays no important role at this point: our results
\cite{AAG2} and those of AAG \cite{GZ2} are different already before averaging.

The difference between AAG's and our
approaches can be illustrated by the example of a
quantum particle (with mass $m$ and coordinate $q$)
interacting with a Caldeira-Leggett
(CL) bath of oscillators (with coupling strength
$\gamma$ and high-frequency cutoff $\omega_c$).
Its reduced density matrix $\rho (q_1,q_2)$ is
also determined by a path integral of the type (\ref{1}).
An expansion in the interaction  $iS_R+S_I$
in the long-time limit yields Fermi's Golden rule with conservation of
the bare energy.
Within this approximation, a particle with initial energy $E \to 0$ at $T=0$
cannot exchange energy and, hence,
phase coherence is preserved. This is the qualitative
argument of Ref. \onlinecite{AAG2}, which refers to a situation
where the total energy of the system is fixed to be the
sum of energies of the noninteracting particles.
On the other hand, if one {\it does not} expand but evaluates the path integral
(\ref{1}) exactly, one obtains \cite{CL} in the long-time limit,
independent of the initial conditions,
$\rho (q_1,q_2) \propto \exp [-m\langle E \rangle (q_1-q_2)^2]$.
Here, the
expectation value of the kinetic energy of the interacting  particle
at $T=0$ is $\langle E\rangle = \gamma\ln(\omega_c/\gamma)>0$.
These results can also be derived by an exact diagonalization
of the initial model~\cite{HA};
however, they cannot be obtained from the Golden rule approach.
The off-diagonal elements of the density matrix decay on a length scale
$\sim 1/\sqrt{m\langle E\rangle }$. Due to interaction this
length scale is finite even at $T=0$; it would diverge if one would assume
$\langle E \rangle \sim T$.
The latter applies to the (obviously coherent) eigenmodes, but
physical quantities which are expressed in the basis $q$ will be
sensitive to the decay of the density matrix $\rho(q_1,q_2)$ as a
function of $q_1-q_2$.
We encountered a similar situation while describing the interacting electron
system. A more detailed discussion is given in Appendix B.

Concerning experiments we can say that our results are in a quantitative
agreement with various experiments, especially in the quasi-1-dimensional case.
The disagreement reported by AAG is observed in strongly disordered
systems with short mean free paths down to several or even one (!) Angstrom.
Such systems are well beyond the applicability of our quasiclassical
theory. An extended discussion of the experiments is given in Appendix C.

We are grateful to our numerous colleagues and friends for support.

\vspace{2mm}

\noindent D.S. Golubev$^{a,c}$ and A.D. Zaikin$^{b,c}$
\vspace{1mm}

\noindent $^a$ Physics Department, Chalmers University,
S-41296 G\"oteborg, Sweden\\
\noindent $^b$ Institut f\"{u}r Theoretische Festk\"orperphysik,
Universit\"at Karlsruhe, 76128 Karlsruhe, FRG\\
\noindent $^c$ Lebedev Physics Institute, 117924 Moscow, Russia

\newpage

\appendix

\section{}

Here we will compare and analyze the expressions for the
weak localization correction to the conductivity in the presence
of interaction obtained in our paper \cite{GZ2} and in Ref. \onlinecite{AAG2}.
The authors \cite{AAG2} raised an extensive critique of our results
which includes qualitative arguments, ``highlighting errors'' in our
calculation and comparison with experiment. The existence of an explicit
error in our calculation would be the most serious argument of AAG
against our theory. Therefore our first task will be to demonstrate
that the AAG statement about the mistakes in our calculation is not correct.
Below we will demonstrate that on a perturbative level our result
\cite{GZ2} is equivalent to that of Ref. \onlinecite{AAG2}.

Before coming to a more technical part of our analysis it would be useful
to briefly remind the reader about the main steps of our calculation \cite{GZ2}.
An attempt to examine our derivation ``step by step'' has been already made
in Sec. 6 of Ref. \onlinecite{AAG2}. Unfortunately the information
contained there is incomplete. In fact from \cite{AAG2} the reader
might get an impression that what we do essentially boils down to two
trivial technical steps (performing the Hubbard-Stratonovich transformation
and representing the Green functions in terms of the path integral) and one
(``incorrect'') disorder average. The actual procedure \cite{GZ2} is
richer both formally and physically.

Our analysis \cite{GZ2} consists of the two main steps.

\begin{enumerate}

\item  The first step
is to reformulate the initial many-body problem with interaction in terms
of a single quantum particle interacting with an effective quantum
environment. Here we indeed use the Hubbard-Stratonovich transformation, but
it is only a simple technical tool. A somewhat less trivial task is to
derive a formally {\it exact} equation of motion for the single electron
density matrix $\rho$ in the presence of interaction (eqs. (24-25) of
Ref. \onlinecite{GZ2}):
\begin{equation}
{\rm i}\frac{\partial\rho_V}{\partial t}=
[H_0-eV_+,\rho_V] - (1-\rho_V)\frac{eV_-}{2}\rho_V -
\rho_V\frac{eV_-}{2}(1-\rho_V),
\label{rho5}
\end{equation}
The matrix $\rho$ is obtained after averaging of $\rho_V$
over the fluctuating fields $V_{\pm}$ carried out with the (again formally
exact) effective action $S[V_+,V_-]$ derived by integrating out
electronic degrees of freedom. No approximations have been made
so far.

Although the expression for $S[V_+,V_-]$ is too complicated to deal with,
some important observations can be made already at this stage. Namely,
the fluctuating field $V_+$ enters the equations just like an external
field whereas the field $V_-$ enters in a qualitatively different manner.
Fluctuations of the field $V_+$ are essentially responsible for dephasing.

In order to proceed further we make the first approximation: we
evaluate the effective action $S[V_+,V_-]$ within RPA.
The action $S$ is now quadratic in $V_{\pm}$ and contains the dielectric
susceptibility $\epsilon (\omega ,k)$ of the effective environment. After
that we easily integrate out the fields $V_{\pm}$ and arrive at the
influence functional $F$ for interacting electrons in a disordered metal.
This completes the first part of our analysis.
We can only add that this approach can be also used in the situations when
approximations other than RPA are more appropriate. In those cases the
action $S[V_+,V_-]$ and the influence functional should be modified
accordingly.

\item As a result of our derivation we arrived at the problem of a
quantum particle in a random potential in the presence of the effective
environment described by the influence functional $F$. The Fermi statistics
and the Pauli principle are explicitely accounted for in the expression for
$F$. The kinetic energy of a particle $E$ is counted from the Fermi
energy $\mu$ and
the states with $E<0$ are forbidden. The second step of our analysis is
to investigate the quantum dynamics of such a particle. According to
the general principles \cite{FV,FH} at this stage the actual physical nature
of the environment already plays no important role, any environment yields
the same effect as long as it is described by the same influence functional.
One can also develop better qualitative understanding of the phenomenon
with the aid of simplier models for the environment. In this sense
the experience gained in the Caldeira-Leggett-type of models
(see e.g. \cite{CL,HA,Leg,Gra,W}) is of particular
interest and will be used in the Appendix B. Now we only mention one
important feature of all these problems: the results for physically
measurable equilibrium quantities do depend on the high frequency
cutoff $\omega_c$ of the effective environment. This dependence has
nothing to do with the excitation of the environment oscillators,
it exists even if the expectation values are calculated in the
{\it true ground state} of the whole system. We will return to this point
further below.

\end{enumerate}

Let us now come to a more technical part of this Appendix. We are going to
test our expression for the influence functional at the level
of the Golden-rule-type perturbation theory employed by the authors \cite{AAG2}.
Since at the first stage of our calculation we only used RPA
(the same approximation was used in \cite{AAG2}) we have to recover all
the same diagrams as in Ref. \onlinecite{AAG2}. However, AAG state that it is not
the case.

We proceed in two steps. We first transform the AAG result for the conductance
and demonstrate that by virtue of the causality principle one
can completely remove the terms of the type $G^RG^AG^RG^A$. We will arrive
at the Eqs. (\ref{dsig}, \ref{dsigma1}) which are exactly equivalent to the
result \cite{AAG2}. Our second step is to expand our expression for the
conductance correction \cite{GZ2} in the interaction terms. This will lead
us to the Eq. (\ref{dsigma00}) which is identical to (\ref{dsigma1}).
The reader not interested in technical details can skip the technical
part and continue reading after the Eq. (\ref{dsigma00}).

We start from reproducing the AAG
expression for the correction to the conductivity due to electron-electron
interaction \cite{AAG2} which they split into two terms
$\delta \sigma _{\alpha \beta }=\delta \sigma _{\alpha \beta }^{\text{deph}}
+\delta \sigma _{\alpha \beta }^{\text{int}}$, where
\begin{eqnarray}
\delta \sigma _{\alpha \beta }^{\text{deph}} &=&-\frac i{16}\int \frac{d
\bbox{r}_1d\bbox{r}_2d\bbox{r}_3d\bbox{r}_4}{{\cal V}}\int \frac{d\omega }{
2\pi }\frac{d\epsilon }{2\pi }\left( \frac d{d\epsilon }\tanh \frac \epsilon
{2T}\right) \left( \coth \frac \omega {2T}+\tanh \frac{\epsilon -\omega }{2T}
\right) \left[ {\cal L}_{34}^R(\omega )-{\cal L}_{34}^A(\omega )\right]
\times   \nonumber \\
&&\left\{ 2\hat j_\alpha \left[ G_{12}^R(\epsilon )-G_{12}^A(\epsilon
)\right] \hat j_\beta \left[ G_{23}^A(\epsilon )G_{34}^A(\epsilon -\omega
)G_{41}^A(\epsilon )-G_{23}^R(\epsilon )G_{34}^R(\epsilon -\omega
)G_{41}^R(\epsilon )\right] +...+\alpha \leftrightarrow \beta \right\} ,
\label{deph}
\end{eqnarray}
\begin{eqnarray}
\delta \sigma _{\alpha \beta }^{\text{int}} &=&-\frac i8\int \frac{d\bbox{r}
_1d\bbox{r}_2d\bbox{r}_3d\bbox{r}_4}{{\cal V}}\int \frac{d\omega }{2\pi }
\frac{d\epsilon }{2\pi }\left( \frac d{d\epsilon }\tanh \frac \epsilon {2T}
\right) \tanh \frac{\epsilon -\omega }{2T}\times   \nonumber \\
&&\big\{ \hat j_\alpha \left[ G_{12}^R(\epsilon )-G_{12}^A(\epsilon
)\right] \hat j_\beta \left[ G_{23}^A(\epsilon )G_{41}^A(\epsilon
)-G_{23}^R(\epsilon )G_{41}^R(\epsilon )\right] \left[ G_{34}^R(\epsilon
-\omega ){\cal L}_{34}^A(\omega )-G_{34}^A(\epsilon -\omega ){\cal L}
_{34}^R(\omega )\right]
\nonumber\\
&& +...+\alpha \leftrightarrow \beta \big\}
\label{int}
\end{eqnarray}
For simplicity we keep the same notations as in \cite{AAG2}: $G^{R(A)}$
are the retarded (advanced) Green functions for noninteracting electrons
and ${\cal L}^{R(A)}$ are photon propagators.
Here ... stands for the terms containing two $\omega $-dependent Green
functions. Such terms do not contribute to $1/\tau{_\varphi}$ neither in
\cite{AAG2} nor in our analysis \cite{GZ2} and therefore these terms will be
ignored further below. For our purposes it will be
convenient to rewrite (\ref{deph},\ref{int}) in the form
\begin{eqnarray}
\delta \sigma _{\alpha \beta } &=&-\frac i{16}\int \frac{d\bbox{r}_1d\bbox{r}
_2d\bbox{r}_3d\bbox{r}_4}{{\cal V}}\int \frac{d\omega }{2\pi }\frac{
d\epsilon }{2\pi }\left( \frac d{d\epsilon }\tanh \frac \epsilon {2T}\right)
\coth \frac \omega {2T}\left[ {\cal L}_{34}^R(\omega )-{\cal L}
_{34}^A(\omega )\right] \times   \nonumber \\
&&\left\{ 2\hat j_\alpha \left[ G_{12}^R(\epsilon )-G_{12}^A(\epsilon
)\right] \hat j_\beta \left[ G_{23}^A(\epsilon )G_{34}^A(\epsilon -\omega
)G_{41}^A(\epsilon )-G_{23}^R(\epsilon )G_{34}^R(\epsilon -\omega
)G_{41}^R(\epsilon )\right] +...+\alpha \leftrightarrow \beta \right\}
\nonumber \\
&&-\frac i{16}\int \frac{d\bbox{r}_1d\bbox{r}_2d\bbox{r}_3d\bbox{r}_4}{{\cal
V}}\int \frac{d\omega }{2\pi }\frac{d\epsilon }{2\pi }\left( \frac d{
d\epsilon }\tanh \frac \epsilon {2T}\right) \tanh \frac{\epsilon -\omega }{2T
}\times   \nonumber \\
&&\big\{ 2\hat j_\alpha \left[ G_{12}^R(\epsilon )-G_{12}^A(\epsilon
)\right] \hat j_\beta \left[ G_{23}^A(\epsilon )G_{41}^A(\epsilon ){\cal L}
_{34}^A(\omega )-G_{23}^R(\epsilon )G_{41}^R(\epsilon ){\cal L}
_{34}^R(\omega )\right] \left[ G_{34}^R(\epsilon -\omega )-G_{34}^A(\epsilon
-\omega )\right]
\nonumber\\
&& +...+\alpha \leftrightarrow \beta \big\} .
\label{dsigma}
\end{eqnarray}
We observe that the factor $\tanh \frac{\epsilon -\omega }{2T}$
enters in this expresion together with the difference $\left[
G_{34}^R(\epsilon -\omega )-G_{34}^A(\epsilon -\omega )\right] $. This
combination is just the Keldysh function
\begin{equation}
G^K(\epsilon ,\bbox{r}_1,\bbox{r}_2)=\tanh \frac \epsilon {2T}\left[
G^R(\epsilon ,\bbox{r}_1,\bbox{r}_2)-G^A(\epsilon ,\bbox{r}_1,\bbox{r}
_2)\right] =\tanh \frac \epsilon {2T}\left[ \frac 1{\epsilon +\mu -\hat H+i0}
-\frac 1{\epsilon +\mu -\hat H-i0}\right] .  \label{GK1}
\end{equation}
This function can also be rewritten as follows
\begin{eqnarray}
G^K(\epsilon ,\bbox{r}_1,\bbox{r}_2) &=&\tanh \frac \epsilon {2T}
\sum\limits_\lambda \left[ \frac 1{\epsilon -\xi _\lambda +i0}-\frac 1{
\epsilon -\xi _\lambda -i0}\right] \Psi _\lambda (\bbox{r}_1)\Psi _\lambda
^{*}(\bbox{r}_2)  \nonumber \\
&=&\tanh \frac \epsilon {2T}\sum\limits_\lambda (-2\pi i)\delta (\epsilon
-\xi _\lambda )\Psi _\lambda (\bbox{r}_1)\Psi _\lambda ^{*}(\bbox{r}_2)
\nonumber \\
&=&\sum\limits_\lambda (-2\pi i)\left( \tanh \frac{\xi _\lambda }{2T}\right)
\delta (\epsilon -\xi _\lambda )\Psi _\lambda (\bbox{r}_1)\Psi _\lambda ^{*}(
\bbox{r}_2)  \nonumber   \\
&=&\sum\limits_\lambda \left( \tanh \frac{\xi _\lambda }{2T}\right) \left[
\frac 1{\epsilon -\xi _\lambda +i0}-\frac 1{\epsilon -\xi _\lambda -i0}
\right] \Psi _\lambda (\bbox{r}_1)\Psi _\lambda ^{*}(\bbox{r}_2)  \nonumber
\\
&=&\int d\bbox{r}^{\prime }\left[ G^R(\epsilon ,\bbox{r}_1,\bbox{r}^{\prime
})-G^A(\epsilon ,\bbox{r}_1,\bbox{r}^{\prime })\right] (\delta (\bbox{r}
^{\prime }-\bbox{r}_2)-2\rho (\bbox{r}^{\prime },\bbox{r}_2)),
\label{trans1}
\end{eqnarray}
where $\xi _\lambda $, $\Psi _\lambda $ are respectively the eigenvalues
and the eigenfunctions of the Hamiltonian $\hat H-\mu $;
$\rho (\bbox{r}^{\prime },
\bbox{r}_2)$ is the equilibrium single electron density matrix, $\hat \rho
=1/(\exp ((\hat H-\mu )/T)+1)$. In a similar manner one obtains
\begin{equation}
\left( \frac d{d\epsilon }\tanh \frac \epsilon {2T}\right) \left[
G^R(\epsilon ,\bbox{r}_1,\bbox{r}_2)-G^A(\epsilon ,\bbox{r}_1,\bbox{r}
_2)\right] =2\int d\bbox{r}^{\prime }\frac{\partial \rho (\bbox{r}_1,\bbox{r}
^{\prime })}{\partial \mu }\left[ G^R(\epsilon ,\bbox{r}^{\prime },\bbox{r}
_2)-G^A(\epsilon ,\bbox{r}^{\prime },\bbox{r}_2)\right] .  \label{trans2}
\end{equation}
We also introduce the evolution operator $\hat U
(t)=\exp (-i(\hat H-\mu)t)$ which is defined both for positive and
negative times. The functions $G^R$ and $G^A$ are related to this
operator by means of the following equations:
\begin{equation}
G^R(t,\bbox{r}_1,\bbox{r}_2)=-i\theta (t)U(t,\bbox{r}_1,\bbox{r}_2);\quad
G^A(t,\bbox{r}_1,\bbox{r}_2)=i\theta (-t)U(t,\bbox{r}_1,\bbox{r}_2).
\label{GU}
\end{equation}
Now let us write down the two equivalent forms of the Keldysh Green
function in the real time representation. We find from (\ref{GK1}):
\begin{eqnarray}
G^K(t,\bbox{r}_1,\bbox{r}_2) &=&\int\limits_{-\infty }^{+\infty }dt^{\prime }
\frac{-iT}{\sinh (\pi T(t-t^{\prime }))}\left[ G^R(t^{\prime },\bbox{r}_1,
\bbox{r}_2)-G^A(t^{\prime },\bbox{r}_1,\bbox{r}_2)\right]   \nonumber \\
&=&-\int\limits_{-\infty }^{+\infty }dt^{\prime }\frac T{\sinh (\pi
T(t-t^{\prime }))}U(t^{\prime },\bbox{r}_1,\bbox{r}_2),  \label{GKA}
\end{eqnarray}
and from (\ref{trans1}) we get
\begin{eqnarray}
G^K(t,\bbox{r}_1,\bbox{r}_2) &=&\int d\bbox{r}^{\prime }\left[ G^R(t,\bbox{r}
_1,\bbox{r}^{\prime })-G^A(t,\bbox{r}_1,\bbox{r}^{\prime })\right] (\delta (
\bbox{r}^{\prime }-\bbox{r}_2)-2\rho (\bbox{r}^{\prime },\bbox{r}_2))
\nonumber \\
&=&-i\int d\bbox{r}^{\prime }U(t,\bbox{r}_1,\bbox{r}^{\prime})
(\delta (\bbox{r}
^{\prime }-\bbox{r}_2)-2\rho (\bbox{r}^{\prime },\bbox{r}_2)).  \label{GKour}
\end{eqnarray}
Analogously we obtain
\begin{eqnarray}
\left( \frac d{d\epsilon }\tanh \frac \epsilon {2T}\right) \left[
G^R(\epsilon ,\bbox{r}_1,\bbox{r}_2)-G^A(\epsilon ,\bbox{r}_1,\bbox{r}
_2)\right]  &\Rightarrow &\int\limits_{-\infty }^{+\infty }dt^{\prime }\frac{
T(t-t^{\prime })}{\sinh (\pi T(t-t^{\prime }))}\left[ G^R(t^{\prime },
\bbox{r}_1,\bbox{r}_2)-G^A(t^{\prime },\bbox{r}_1,\bbox{r}_2)\right]
\nonumber \\
&=&\int\limits_{-\infty }^{+\infty }dt^{\prime }\frac{-iT(t-t^{\prime })}{
\sinh (\pi T(t-t^{\prime }))}U(t^{\prime },\bbox{r}_1,\bbox{r}_2);
\label{dtanhA}
\end{eqnarray}
and
\begin{eqnarray}
2\int d\bbox{r}^{\prime }\frac{\partial \rho (\bbox{r}_1,\bbox{r}^{\prime })
}{\partial \mu }\left[ G^R(\epsilon ,\bbox{r}^{\prime },\bbox{r}
_2)-G^A(\epsilon ,\bbox{r}^{\prime },\bbox{r}_2)\right]  &\Rightarrow &2\int
d\bbox{r}^{\prime }\frac{\partial \rho (\bbox{r}_1,\bbox{r}^{\prime })}{
\partial \mu }\left[ G^R(t,\bbox{r}^{\prime },\bbox{r}_2)-G^A(t,\bbox{r}
^{\prime },\bbox{r}_2)\right]   \nonumber  \\
&=&-2i\int d\bbox{r}^{\prime }\frac{\partial \rho (\bbox{r}_1,\bbox{r}
^{\prime })}{\partial \mu }U(t,\bbox{r}^{\prime },\bbox{r}_2).
\label{dtanhour}
\end{eqnarray}
It is easy to observe that the eqs. (\ref{GKA},\ref{dtanhA}) contain
the integral over time which does not enter the eqs.
(\ref{GKour},\ref{dtanhour}). It is this additional time integration
that leads to violation of the normal time ordering at the level of the
perturbation theory and is responsible for the appearence of the diagrams
$G^RG^AG^RG^A$. The interpretation of such diagrams in terms of the path
integral is not possible. However, if one uses the other form of the same
functions (\ref{GKour},\ref{dtanhour}) the normal time ordering is
automatically restored, the combinations $G^RG^AG^RG^A$ dissapear
due to the causality principle
and the path integral interpretation of the remaining terms of the
perturbation theory can be made.

We emphasize that all the above transformations are exact and have the
advantage that in the final expressions only the propagators depend
on the frequencies $\epsilon $ and $\omega $ (except for the factor $\coth \frac
\omega {2T}$ in $\delta \sigma _{\alpha \beta }$). This allows one to use
the analytical properties of the propagators related to the causality principle.
Namely, $G^R(\epsilon )$ and ${\cal L}^R(\omega )$ have no singularities in
the upper half-plane, while $G^A(\epsilon )$ and ${\cal L}^A(\omega )$ are
analytic functions in the lower half-plane. Making use of these properties
one can easily prove the identities
\begin{eqnarray}
\int d\omega {\cal L}^R(\omega )G^A(\epsilon -\omega ) &\equiv &0,\int
d\epsilon G_{12}^A(\epsilon )G_{23}^A(\epsilon )G_{34}^A(\epsilon -\omega
)G_{41}^A(\epsilon )\equiv 0,  \nonumber \\
\int d\omega {\cal L}^A(\omega )G^R(\epsilon -\omega ) &\equiv &0,\int
d\epsilon G_{12}^R(\epsilon )G_{23}^R(\epsilon )G_{34}^R(\epsilon -\omega
)G_{41}^R(\epsilon )\equiv 0.  \label{causality}
\end{eqnarray}
Consider e.g. the integral $\int d\omega {\cal L}^R(\omega
)G^A(\epsilon -\omega )$. Since both functions ${\cal L}
^R(\omega )$ and $G^A(\epsilon -\omega )$ are regular in the upper
half-plane, the integral vanishes. Alternatively, we can write $\int d\omega
{\cal L}^R(\omega )G^A(\epsilon -\omega )=2\pi\int dt\exp (i\epsilon t){\cal L}
^R(t)G^A(t)$ and note that ${\cal L}^R(t)\equiv 0$ for $t<0$ due to
the causality principle, while $G^A(t)\equiv 0$ for $t>0$ and
the integral is identically equal to zero. Analogously
one can prove all the other identities (\ref{causality}).

The corrections to the conductivity can now be considerably simplified:
\begin{eqnarray}
\delta \sigma _{\alpha \beta } &=&-\frac i4\int \frac{d\bbox{r}_1d\bbox{r}_2d
\bbox{r}_3d\bbox{r}_4d\bbox{r}_5}{{\cal V}}\int \frac{d\omega }{2\pi }\frac{
d\epsilon }{2\pi }\coth \frac \omega {2T}\left[ {\cal L}_{34}^R(\omega )-
{\cal L}_{34}^A(\omega )\right] \times   \nonumber \\
&&\left\{ \hat j_\alpha \left[ G_{15}^R(\epsilon )\frac{\partial \rho _{52}}{
\partial \mu }\right] \hat j_\beta G_{23}^A(\epsilon )G_{34}^A(\epsilon
-\omega )G_{41}^A(\epsilon )+\hat j_\alpha \left[ G_{15}^A(\epsilon )\frac{
\partial \rho _{52}}{\partial \mu }\right] \hat j_\beta G_{23}^R(\epsilon
)G_{34}^R(\epsilon -\omega )G_{41}^R(\epsilon )+..+\alpha \leftrightarrow
\beta \right\}   \nonumber \\
&&-\frac i4\int \frac{d\bbox{r}_1d\bbox{r}_2d\bbox{r}_3d\bbox{r}_4d\bbox{r}
_5d\bbox{r}_6}{{\cal V}}\int \frac{d\omega }{2\pi }\frac{d\epsilon }{2\pi }
\left\{ -\hat j_\alpha G_{15}^R(\epsilon )\frac{\partial \rho _{52}}{
\partial \mu }\hat j_\beta G_{23}^A(\epsilon )
\left[G_{36}^A(\epsilon -\omega
)\left( 1-2\rho \right) _{64}\right]
G_{41}^A(\epsilon ){\cal L}_{34}^A(\omega
)\right.   \nonumber \\
&&+\left. \hat j_\alpha G_{15}^A(\epsilon )\frac{\partial \rho _{52}}{
\partial \mu }\hat j_\beta G_{23}^R(\epsilon )
\left[G_{36}^R(\epsilon -\omega
)\left( 1-2\rho \right) _{64}\right]
G_{41}^R(\epsilon ){\cal L}_{34}^R(\omega
)+...+\alpha \leftrightarrow \beta \right\} .  \label{dsig}
\end{eqnarray}
We observe that the terms of the type $\hat j
_\alpha G_{12}^R(\epsilon )\hat j_\beta G_{23}^A(\epsilon )G_{34}^R(\epsilon
-\omega )G_{41}^A(\epsilon )$ do not enter the expression (\ref{dsig}) at all.
For later purposes it will be useful to rewrite the above expression in the form
of the time integral:
\begin{eqnarray}
\delta \sigma _{\alpha \beta } &=&-\frac{e^2}2\int \frac{d\bbox{r}
_1d\bbox{r}_2d\bbox{r}_3d\bbox{r}_4d\bbox{r}_5}{{\cal V}}\int\limits_0^{+
\infty }dt_1\int\limits_0^{t_1}dt_2\int\limits_0^{t_2}dt_3\times   \nonumber
\\
&&\left\{ \hat j_\alpha U_{15}(t_1)\frac{\partial \rho _{52}}{\partial \mu }
\hat j_\beta U_{23}^{+}(t_3)\left[ I_{34}(t_2-t_3)U_{34}^{+}(t_2-t_3)\right]
U_{41}^{+}(t_1-t_2)\right.   \nonumber \\
&&+\left. \hat j_\alpha U_{15}^{+}(t_1)\frac{\partial \rho _{52}}{\partial
\mu }\hat j_\beta U_{23}(t_3)\left[ I_{34}(t_2-t_3)U_{34}(t_2-t_3)\right]
U_{41}(t_1-t_2)+...+\alpha \leftrightarrow \beta \right\}   \nonumber \\
&&-\frac{ie^2}4\int \frac{d\bbox{r}_1d\bbox{r}_2d\bbox{r}_3d\bbox{r}_4d
\bbox{r}_5d\bbox{r}_6}{{\cal V}}\int\limits_0^{+\infty
}dt_1\int\limits_0^{t_1}dt_2\int\limits_0^{t_2}dt_3\times   \nonumber \\
&&\left\{ -\hat j_\alpha U_{15}(t_1)\frac{\partial \rho _{52}}{\partial \mu }
\hat j_\beta U_{23}^{+}(t_3)\left[ R_{34}(t_2-t_3)U_{36}^{+}(t_2-t_3)\left(
1-2\rho \right) _{64}\right] U_{41}^{+}(t_1-t_2)\right.   \nonumber \\
&&+\left. \hat j_\alpha U_{15}^{+}(t_1)\frac{\partial \rho _{52}}{\partial
\mu }\hat j_\beta U_{23}(t_3)\left[ R_{34}(t_2-t_3)U_{36}(t_2-t_3)\left(
1-2\rho \right) _{64}\right] U_{41}(t_1-t_2)+...+\alpha \leftrightarrow
\beta \right\} ;  \label{dsigma1}
\end{eqnarray}
where
\begin{eqnarray*}
R(t,\bbox{r}) &=&\int \frac{d\omega d^3k}{(2\pi )^4}\frac{4\pi }{k^2\epsilon
(\omega ,k)}e^{-i\omega t+i\bbox{kr}}=-\frac 1{e^2}{\cal L}^R(t,\bbox{r})=-\frac 1{
e^2}{\cal L}^A(-t,\bbox{r}), \\
I(t,\bbox{r}) &=&\int \frac{d\omega d^3k}{(2\pi )^4}\text{Im}\left( \frac{
-4\pi }{k^2\epsilon (\omega ,k)}\right) \coth \left( \frac \omega {2T}
\right) e^{-i\omega t+i\bbox{kr}}
=\frac 1{2e^2i}\int \frac{d\omega d^3k}{
(2\pi )^4}\coth \left( \frac \omega {2T}\right) \left[ {\cal L}^R(\omega ,k)-
{\cal L}^A(\omega ,k)\right] e^{-i\omega t+i\bbox{kr}}.
\end{eqnarray*}

Now we will demonstrate that the equation (\ref{dsigma1}) can be obtained
within the path integral formalism.
The formal expression for the conductivity has the form \cite{GZ2}
\begin{equation}
\sigma=
\frac{e^2}{3m}
\int\limits_{-\infty}^t dt'\int d\bbox{r}_{i1}d\bbox{r}_{i2}
\left.\left(\nabla_{r_{1f}}-\nabla{r_{2f}}\right)\right|_{\bbox{r}_{1f}
=\bbox{r_{2f}}}J(t,t';\bbox{r}_{1f},\bbox{r}_{2f};\bbox{r}_{1i},\bbox{r}_{2i})
(\bbox{r}_{1i}-\bbox{r}_{2i})\rho_0(\bbox{r}_{1i},\bbox{r}_{2i}).
\label{sigma00}
\end{equation}
The kernel $J$ is given by the path integral over electron coordinates
and momentums $\bbox{r}_1(t),\bbox{p}_1(t)$ and $\bbox{r}_2(t),\bbox{p}_2(t)$
corresponding respectively to the forward and backward parts of the Keldysh
contour. The explicit expression for this kernel reads \cite{GZ2}:
\begin{eqnarray}
J(t,t';\bbox{r}_{1f},\bbox{r}_{2f};\bbox{r}_{1i},\bbox{r}_{2i})&=&
\int\limits_{\bbox{r}_1(t')=\bbox{r}_{1i}}^{\bbox{r}_1(t)=\bbox{r}_{1f}}
{\cal D}\bbox{r}_1\int\limits_{\bbox{r}_2(t')=\bbox{r}_{2i}}^{\bbox{r}_2(t)
=\bbox{r}_{2f}}{\cal D}\bbox{r}_2\int{\cal D}\bbox{p}_1\int{\cal D}
\bbox{p}_2\times
\nonumber \\
&&
\times \exp\big\{iS_0[\bbox{r}_1,\bbox{p}_1]-iS_0[\bbox{r}_2,\bbox{p}_2]-
iS_R[\bbox{r}_1,\bbox{p}_1,\bbox{r}_2,\bbox{p}_2]-
S_I[\bbox{r}_1,\bbox{r}_2]\big\};
\label{J17}
\end{eqnarray}
where
\begin{equation}
S_0[\bbox{r},\bbox{p}]=\int\limits_{t'}^t dt''
\bigg(\bbox{p\dot r} - \frac{\bbox{p}^2}{2m} - U(\bbox{r})\bigg);
\label{S_0}
\end{equation}
\begin{eqnarray}
S_R[\bbox{r}_1,\bbox{p}_1,\bbox{r}_2,\bbox{p}_2]&=&
\frac{e^2}{2}\int\limits_{t'}^t dt_1 \int\limits_{t'}^t dt_2
\big\{R(t_1-t_2,\bbox{r}_1(t_1)-\bbox{r}_1(t_2))
\big[1-2n\big(\bbox{p}_1(t_2),\bbox{r}_1(t_2)\big)\big]
\nonumber \\
&&
-R(t_1-t_2,\bbox{r}_2(t_1)-\bbox{r}_2(t_2))
\big[1-2n\big(\bbox{p}_2(t_2),\bbox{r}_2(t_2)\big)\big]
\nonumber \\
&&
+R(t_1-t_2,\bbox{r}_1(t_1)-\bbox{r}_2(t_2))
\big[1-2n\big(\bbox{p}_2(t_2),\bbox{r}_2(t_2)\big)\big]
\nonumber \\
&&
-R(t_1-t_2,\bbox{r}_2(t_1)-\bbox{r}_1(t_2))
\big[1-2n\big(\bbox{p}_1(t_2),\bbox{r}_1(t_2)\big)\big]
\big\};
\label{SR}
\end{eqnarray}
and
\begin{eqnarray}
S_I[\bbox{r}_1,\bbox{r}_2]&=&
\frac{e^2}{2}\int\limits_{t'}^t dt_1 \int\limits_{t'}^t dt_2
\bigg\{I(t_1-t_2,\bbox{r}_1(t_1)-\bbox{r}_1(t_2))+
I(t_1-t_2,\bbox{r}_2(t_1)-\bbox{r}_2(t_2))
\nonumber \\
&&
-I(t_1-t_2,\bbox{r}_1(t_1)-\bbox{r}_2(t_2))-
I(t_1-t_2,\bbox{r}_2(t_1)-\bbox{r}_1(t_2))\bigg\}.
\label{SI}
\end{eqnarray}
The functions $R(t,\bbox{r})$ and $I(t,\bbox{r})$ were already defined above.

In order to obtain the perturbative result (\ref{dsigma1}) from the
formally exact
expression (\ref{sigma00}) one needs to expand the kernel $J$ (\ref{J17})
in $iS_R+S_I$. In the first
order one obtains eight different terms. Again we will consider only the terms
contributing to $1/\tau_\varphi$, i.e. the terms containing
$R(t_1-t_2,\bbox{r}_1(t_1)-\bbox{r}_1(t_2)),$
$R(t_1-t_2,\bbox{r}_2(t_1)-\bbox{r}_2(t_2)),$
$I(t_1-t_2,\bbox{r}_1(t_1)-\bbox{r}_1(t_2))$ and
$I(t_1-t_2,\bbox{r}_2(t_1)-\bbox{r}_2(t_2)).$
Diagrammatically, these contributions are described by the first two diagrams
shown in Fig.4 of Ref. \onlinecite{AAG2}.
Four other terms which relate two different branches of the Keldysh contour
and contain both $\bbox{r}_1$ and $\bbox{r}_2,$ e.g.  of the type
$I(t_1-t_2,\bbox{r}_1(t_1)-\bbox{r}_2(t_2)),$ are described by the last diagram
in Fig. 4 of Ref. \onlinecite{AAG2}. These terms give the contributions
containing two Green functions depending on the frequency $\omega$,
and were denoted as ...  in (\ref{dsigma1}). Such terms are fully reproduced
within our method as well, however we will not consider them here for the
sake of simplicity.

The correction to the kernel $J$ due to the term
$I(t_1-t_2,\bbox{r}_1(t_1)-\bbox{r}_1(t_2))$ has the form
\begin{eqnarray}
\delta
J_I^{11}(t,t';\bbox{r}_{1f},\bbox{r}_{2f};\bbox{r}_{1f},\bbox{r}_{2f})
&=&-e^2\int\limits_{t'}^{t} dt_3 \int\limits_{t'}^{t_3} dt_2
\int\limits_{\bbox{r}_1(t')=\bbox{r}_{1i}}^{\bbox{r}_1(t)=\bbox{r}_{1f}}
{\cal D}\bbox{r}_1\int\limits_{\bbox{r}_2(t')=\bbox{r}_{2i}}^{\bbox{r}_2(t)
=\bbox{r}_{2f}}{\cal D}\bbox{r}_2\int{\cal D}\bbox{p}_1\int{\cal D}
\bbox{p}_2\times
\nonumber \\
&&\times I(t_3-t_2,\bbox{r}_1(t_3)-\bbox{r}_1(t_2))
\exp\big\{ iS_0[\bbox{r}_1,\bbox{p}_1]-iS_0[\bbox{r}_2,\bbox{p}_2]\big\}
\nonumber \\
&=&-e^2\int d\bbox{r}_3 \int d\bbox{r}_4
\int\limits_{t'}^{t} dt_3 \int\limits_{t'}^{t_3} dt_2\times
\nonumber\\
&&U^{+}_{\bbox{r}_{2f},\bbox{r}_{2i}}(t-t')
U_{\bbox{r}_{1f},3}(t-t_3)I_{34}(t_3-t_2)U_{34}(t_3-t_2)
U_{4,\bbox{r}_{1i}}(t_2-t').
\label{dJI11}
\end{eqnarray}
Here we made use of a simple property of a path integral:
\begin{equation}
\int\limits_{\bbox{r}(t')=\bbox{r}_{i}}^{\bbox{r}(t)=\bbox{r}_{f}}
{\cal D}\bbox{r}\int{\cal D}\bbox{p} f(t'',\bbox{r}(t''))
\exp\big\{iS_0[\bbox{r},\bbox{p}]\big\}=
\int d\bbox{r}'' U(t-t'';\bbox{r}_f,\bbox{r}'')f(t'',\bbox{r}'')
U(t''-t;\bbox{r}'',\bbox{r}_i),
\label{svojstvo}
\end{equation}
which holds for an arbitrary function $f(t'',\bbox{r}(t''))$. Actually in
deriving (\ref{dJI11}) the property (\ref{svojstvo}) was used twice because the
function of two arguments $I(t_2-t_3,\bbox{r}_1(t_2)-\bbox{r}_1(t_3))$ enters
under the integral (\ref{dJI11}).
Already at this stage one can observe the similarity between the expression
(\ref{dJI11}) and the second term in the expression (\ref{dsigma1}). To
establish the equivalence between these two expressions the
following steps are in order:
i) after substituting the result (\ref{dJI11}) into the expression for
the conductivity (\ref{sigma00}) and applying the current
operator $\bbox{j}=(ie/m)(\nabla_{{r}_{1f}}-\nabla_{r_{2f}})$ one puts
$\bbox{r}_{1f}=\bbox{r}_{2f}=\bbox{r}_2$, $\bbox{r}_{1i}=\bbox{r}_1$,
$\bbox{r}_{2i}=\bbox{r}_5$; ii) one denotes $t-t'\to t_1$,
$t-t_2\to t_2$, $t-t_3\to t_3$; iii) one introduces an additional integration
$\int d\bbox{r}_2/{\cal V}$ which is just averaging of
the expression (\ref{sigma00}) over the sample volume and iv) one
transforms the effective initial density matrix as follows
\begin{equation}
(\bbox{r}_{1i}-\bbox{r}_{2i})\rho_0(\bbox{r}_{1i},\bbox{r}_{2i})=
i\sum\limits_{\lambda_1\lambda_2}
\frac{\langle\Psi_{\lambda_1}|\bbox{p}|\Psi_{\lambda_2}\rangle}{m}
\frac{n(\xi_{\lambda_1})-n(\xi_{\lambda_2})}{\xi_{\lambda_1}-\xi_{\lambda_2}}
\Psi_{\lambda_1}(\bbox{r}_{1i})\Psi_{\lambda_2}^*(\bbox{r}_{2i})
\simeq -i\frac{\bbox{\hat p}}{m}
\frac{\partial\rho(\bbox{r}_{1i},\bbox{r}_{2i})}{\partial\mu}.
\label{transform}
\end{equation}
After these transformations one can immediately observe the equivalence
of the results obtained by means of two methods \cite{GZ2} and \cite{AAG2}
on the level of the perturbation theory. The terms arising from the
real part of the action $S_R$
can be transformed analogously, the only difference in this case is the
presence of an additional factor $(1-2\rho)_{64}$ related to the term
$1-2n(\bbox{p},\bbox{r})$ in the expression (\ref{SR}). Finally we get
\begin{eqnarray}
\delta \sigma &=&-\frac{e^3}{3}\int \frac{d\bbox{r}
_1d\bbox{r}_2d\bbox{r}_3d\bbox{r}_4d\bbox{r}_5}{{\cal V}}\int\limits_0^{+
\infty }dt_1\int\limits_0^{t_1}dt_2\int\limits_0^{t_2}dt_3\times   \nonumber
\\
&&\left\{ \frac{\bbox{\hat p}}{m} \frac{\partial \rho _{15}}{\partial \mu
}U_{52}(t_1) \bbox{\hat j} U_{23}^{+}(t_3)\left[
I_{34}(t_2-t_3)U_{34}^{+}(t_2-t_3)\right] U_{41}^{+}(t_1-t_2)\right.
\nonumber  \\
&&+\left. \frac{\bbox{\hat p}}{m} \frac{\partial \rho _{15}}{\partial
\mu }U_{52}^{+}(t_1)
\bbox{\hat j} U_{23}(t_3)\left[ I_{34}(t_2-t_3)U_{34}(t_2-t_3)\right]
U_{41}(t_1-t_2)+... \right\}
\nonumber \\
&&-\frac{ie^3}{6}\int \frac{d\bbox{r}_1d\bbox{r}_2d\bbox{r}_3d\bbox{r}_4d
\bbox{r}_5d\bbox{r}_6}{{\cal V}}\int\limits_0^{+\infty
}dt_1\int\limits_0^{t_1}dt_2\int\limits_0^{t_2}dt_3\times
\nonumber \\
&&\left\{ -\frac{\bbox{\hat p}}{m}
\frac{\partial \rho _{15}}{\partial \mu }U_{52}(t_1)
\bbox{\hat j} U_{23}^{+}(t_3)\left[ R_{34}(t_2-t_3)U_{36}^{+}(t_2-t_3)\left(
1-2\rho \right)_{64}\right] U_{41}^{+}(t_1-t_2)\right.
\nonumber \\
&&+\left. \frac{\bbox{\hat p}}{m}
\frac{\partial \rho_{15}}{\partial
\mu } U_{52}^{+}(t_1)
\bbox{\hat j} U_{23}(t_3)\left[ R_{34}(t_2-t_3)U_{36}(t_2-t_3)\left(
1-2\rho \right)_{64}\right] U_{41}(t_1-t_2)+... \right\}.
\label{dsigma00}
\end{eqnarray}
In order to verify complete equivalence of (\ref{dsigma1}) and (\ref{dsigma00})
one should a) replace the operator $e\bbox{\hat p}/m$ by
$\bbox{\hat j}$; b) adjust the factor 3 by observing that
(\ref{dsigma00}) and (\ref{dsigma1}) are the corrections respectively to
to the scalar conductivity and the conductivity tensor (in the isotropic case
one has $\delta\sigma=(\delta\sigma_{xx}+\delta\sigma_{yy}+\delta\sigma_{zz})/3$)
and c) adjust another factor 2 having in mind symmetrization of (\ref{dsigma1})
with respect to indices $\alpha$ and $\beta$.
Also, one should keep in mind that the operator
$\partial\rho_{15}/\partial\mu$ commutes with the evolution operator
$U_{52}$.
This completes the
proof of equivalence of the results (\ref{dsigma1}) and (\ref{dsigma00}).

Thus the AAG statement that within our analysis ``only some paths were
selected by hand'' is false. As it was explained above the paths presented
e.g. in Fig. 10c of Ref. \onlinecite{AAG2} {\it cannot} appear in
the path integral,
they are forbidden by the causality principle. Unfortunately AAG did not
indicate the direction of the electron motion (e.g. by arrows) in their
Fig. 10c. Otherwise it would be completely clear that at some parts of
this path the electron moves backward in time. Such diagrams can appear
from direct multiplication of the Keldysh matrices, but not in the path
integral.

In any case the technical issue with ``missing diagrams'' is settled and
now we can come to the central question: if there are no calculational errors
what causes the difference between the results of Refs. \onlinecite{AAG2} and
\onlinecite{GZ2}? As it was already discussed AAG proceed perturbatively in the
interaction. They state that we do the same: according to AAG our
procedure ``is nothing but a perturbative expansion'' \cite{AAG1} and our
results are ``purely perturbative'' \cite{AAG2}. If this were true, our
final results could be compared directly indeed. However, this is
{\it not true}. Of course, in some sense we also proceed
perturbatively when we expand the exact effective action $S[V_{\pm}]$ in
powers of $V_{\pm}$ in the exponent. But this is just RPA,
and it has nothing to do with the Golden-rule-type perturbation theory
developed in \cite{AAG2}. In this sense our procedure is essentially
nonperturbative and (within RPA) includes processes in all orders.

On a slightly more formal level we can reformulate the difference as follows.
The conductance (\ref{sigma00}) is determined by the path integral (\ref{J17}).
As it was demonstrated above the AAG perturbative procedure is equivalent
to expanding $J$ up to
the first order in $iS_R+S_I$, integrating over time in the infinite limits
(this yields energy conservation) and averaging the result over disorder
within the quasiclassical approximation. Our procedure {\it does not} involve
the expansion in $iS_R+S_I$. We just evaluate the complete path integral
within the quasiclassical approximation taking into account all saddle
point paths and average over disorder. The latter average
is quite trivial because the part of $S_R$ which could be important for
dephasing disappears already before averaging.

One might think that both procedures should give the same result as long
as the interaction is weak and the terms $iS_R+S_I$ are small. Here, however,
one should be cautious: these terms are defined by nonlocal in time
expressions (\ref{SR},\ref{SI}) and involve integrals over the two times.
These terms are never small at low $T$ as long as time is not small.
Therefore in general one cannot expand and should evaluate the
whole path integral no matter how weak the interaction is.

Thus not only the equivalence of the results of refs. \onlinecite{AAG2} and
\onlinecite{GZ2} on the perturbative level but also an important difference
between these results is established. In both papers the conductance of a
disordered metal (\ref{sigma00}) in the presence of interaction was calculated.
In our paper \cite{GZ2} the {\it full} path integral (\ref{J17}) was
evaluated whereas the analysis \cite{AAG2} is equivalent to the first
order expansion of this path integral in $iS_R+S_I$. It is exactly due
to this reason the results \cite{AAG2} and \cite{GZ2} differ. The
average over disorder emphasized in \cite{AAG2} is not important at this
stage: the results are different already before averaging.

We can also add that for simplicity in Ref. \onlinecite{GZ2} we did not consider
the effect of magnetic field $H$. This effect can be trivially incorporated
into our analysis by adding the term $e\bbox{\dot{r}A}(\bbox{r})$ to the
electron Lagrangian, $\bbox{A}$ is the corresponding vector potential. The
whole procedure \cite{GZ2} remains the same except the action $S_0$ will now
depend on $\bbox{A}$. Averaging over disorder at the last stage of the
calculation will yield the equations (2.42) from \cite{AAG2} which define
the dependence of the magnetoconductance curves on $H$. This part of
the calculation is standard and was discussed in details e.g. in \cite{CS}.
We would only like to emphasize that even in the presence of the magnetic
field $H$ one is not allowed to expand the path integral $J$ (\ref{J17})
in powers of $iS_R+S_I$ for all relevant fields $\tau_H \gg \tau_e$, where
$\tau_H$ is the dephasing time due to the magnetic field, see eqs. (2.9) of
Ref. \onlinecite{AAG2}.

Having established the important formal difference between the procedures
\cite{AAG2} and \cite{GZ2} we could, in principle, already conclude our
consideration. At this stage one does not even need to introduce and discuss
such concepts as ``decoherence'', ``dephasing time'' etc. Just viewing
$\tau_{\varphi}$ as a formal parameter extracted in a standard way from
the magnetoconductance measurements one immediately observes that the
perturbative calculation \cite{AAG2} yields $\tau_{\varphi} \to \infty$
at $T \to 0$, while if one evaluates the whole path integral (\ref{J17})
(which contains all the perturbative terms \cite{AAG2} plus infinitely many
other terms which contribution is not smaller than that kept in
\cite{AAG2}) one
obtains a finite result at low $T$ in agreement with many experiments.
We believe, however, that it is important not only to establish the
formal difference between the two approaches \cite{AAG2} and \cite{GZ2}
but also to understand the physical reasons for this difference to occur.
The corresponding discussion is presented below.

\section{}

Let us first recall the qualitative arguments of AAG concerning the
effect of quantum decoherence \cite{AAG2}: ``Consider
a quantum particle which moves in the environment of harmonic oscillators:
each oscillator is characterized by its frequency $\omega$. The result of
the collision of the particle with an oscillator depends on the relation
between $\hbar \omega$ and the temperature.'' AAG first consider the case
$\hbar \omega \ll T$ and find that in this case ``the probability for
the inelastic collision is substantial''. Then they write: ``The situation
for $\hbar \omega \gg T$ is quite different. Indeed, up to the
exponentially small terms the oscillator is in the ground state and the
particle has the energy smaller than $\hbar \omega$''. Here we already
have two questions: (i) it remains unclear whether the interacting or
noninteracting ground state of the system is meant and (ii) it is also not
clear whether the whole system is in equilibrium at all. But let us first
finish the citation from \cite{AAG2}:
``Therefore, the energy transfer is forbidden by the energy conservation
and the collision is {\it elastic}. Therefore, there is no difference
whatsoever with the collision with such an oscillator and with the quenched
disorder, which definitely does not dephase''.

Here we definitely agree with
the last part of this sentence, namely quenched disorder indeed cannot dephase.
The remaining part emphasizes the {\it main physical difference} between
\cite{AAG2} and our work. The AAG arguments apply if
one considers a scattering problem in which the total energy of the system
is a sum of those for the {\it noninteracting} particle and the oscillators
and it (the energy) is conserved during the whole process.
Then, if initially all the oscillators were in their noninteracting ground
states and the particle energy was small, in the end this particle will have
the same energy because none of the oscillators can be excited. Hence, the
collision is elastic and no dephasing takes place.

We, however, are interested in another physical situation. Namely,
we would like to describe the properties of a quantum particle which is
in equilibrium with an (infinite) bath of oscillators or close to this
equilibrium. In this
case the energy exchange between the particle and the bath is possible and
the total energy of this {\it interacting} system is different (in our case larger)
as compared to a sum of energies of noninteracting particles.
Then for various physical situations one can demonstrate
(see e.g. \cite{CL,HA,Leg,Gra,W}) that interaction of a quantum particle with
other quantum degrees of freedom is qualitatively different from
the effect of a static potential. Since the system with harmonic
oscillators is simple enough one can illustrate the difference between the two
physical situations by means of an exact calculation. This will be done below.

The density matrix of a quantum particle $q$ interacting with other
quantum degrees of freedom can be represented in the form
\begin{equation}
\rho (q_f,q'_f)=
\int dq_idq'_iJ(q_f,q'_f,t;q_i,q'_i,0)\rho_i(q_i,q'_i),
\label{rho1}
\end{equation}
where $\rho_i(q_i,q'_i)$ is the initial density matrix and
\begin{equation}
J=\int_{q_i}^{q_f}{\cal D}q_1\int_{q'_i}^{q'_f}{\cal D}q_2
\exp ({\rm i}S_0[q_1(t)]-{\rm i}S_0[q_2(t)]
-{\rm i}S_R[q_1,q_2]-S_I[q_1,q_2]).
\label{J}
\end{equation}
Here $S_0[q]$ is the action of a noninteracting particle, while the
terms $S_{R/I}$ account for interaction and define the influence functional
for an effective quantum environment \cite{FV,FH}. It is easy to observe
that the Eq. (\ref{J}) has essentially the same form as the Eq. (\ref{J17}).
If the environment consists of harmonic oscillators with frequencies $\omega_n$
and unity masses, the influence functional can be easily found
provided one assumes bilinear in coordinates interaction between the particle
and the oscillators. Defining $q_+=(q_1+q_2)/2$ and $q_-=q_1-q_2$ one readily
obtains
\cite{FV,FH}
\begin{equation}
S_R=\sum_{n}\frac{C^2}{\omega_n}\int_{0}^{t}dt_1dt_2q_-(t_1)
\sin (\omega_n (t_1-t_2))q_+(t_2),
\label{SR0}
\end{equation}
\begin{equation}
S_I=\sum_{n}\frac{C^2}{2\omega_n}\coth \left( \frac{\omega_n}{2T}\right)
\int_{0}^{t}dt_1dt_2q_-(t_1)
\cos (\omega_n (t_1-t_2))q_-(t_2).
\label{SI0}
\end{equation}
where $C$ is a constant which governs the strength of interaction. At
this stage the problem (\ref{J}-\ref{SI0}) is qualitatively analogous to one
we arrive in our analysis of interacting electrons in a disordered metal after
we employ an RPA expansion of the exact effective action (cf. eqs.
(\ref{J17}-\ref{SI})).
In a model with oscillators we deal with the quadratic effective action from
the very beginning.

Let us assume $C$ to be very small and try to find the transition
probability
\begin{equation}
W_{if}= \int dq_f dq'_f dq_i dq'_i\psi_f(q_f)
\psi_f^*(q'_f)J(q_f,q'_f,t;q_i,q'_i,0)\psi_i(q_i)\psi_i^*(q'_i)
\label{Wif}
\end{equation}
from some initial state with the wave function
$\psi_i(q)$ and the energy $E_i$ to some other orthogonal
state $\psi_f(q)$, $E_f$
just by expanding the kernel $J$ (\ref{J}) in powers of $C$. The calculation
is trivial (see e.g. \cite{FH}) and we only reproduce the result:
\begin{equation}
W_{if}={\rm Re}\sum_n \frac{\pi C^2 |\langle i|q|f\rangle |^2}{2\omega_n}
\int dt \int dt' e^{-i(E_f-E_i)(t-t')} \int \frac{d\omega}{2\pi}
e^{-i\omega (t-t')}
\biggl(\coth \frac{\omega}{2T}+1\biggr)[\delta(\omega_n-\omega )-
\delta(\omega_n+\omega )]
\label{Wif2}
\end{equation}
Integration over $t-t'$ in the infinite limits yields the delta-function
$\delta (E_f+\omega -E_i)$ which ensures the energy conservation. At $T\to 0$
only positive $\omega$ contribute and the second delta-function in the square
brackets fails. Thus in this case the transitions are only possible if
$E_f+\omega_n-E_i=0$, i.e. for $E_i\leq E_f$ the transition never happens
because the system has no energy to excite the oscillator $\omega_n$.
Thus it always stays in its initial state, i.e. $W_{if}\equiv 0$
and the quantum coherence is never lost. At nonzero $T$ and
$E_i\leq E_f$ we have $W_{if} \neq 0$, but only oscillators with small
frequencies $\omega_n \lesssim T$ can take part in the transitions, so
that $W_{if}$ is small as long as $T$ is sufficiently low.

The above simple example just illustrates the qualitative arguments of
the authors \cite{AAG2} presented above. Exactly the same physical reasons
are behind the cancellation of diagrams demonstrated in \cite{AAG2}.
This cancellation will always take place
at $T \to 0$ within the Golden-rule-type perturbation theory just
because of energy conservation. The only difference is that instead
of the combination ``$\coth +1$'' for Bose particles there appears the
combination ``$\coth -\tanh$'' in the case of electrons. This is well
known and was also rederived from our formalism (see Sec. 5 of
Ref. \onlinecite{GZ2}).

Thus the AAG calculation \cite{AAG2} is fully
consistent with their qualitative arguments. Both describe the same
physical situation: the total energy of the system ``particle+oscillators''
is equal to its value without interaction and fixed during the whole
process.

It is obvious, however, that the formalism allows to go beyond the simple
Golden-rule-type perturbation theory and to provide a full description
of an interacting system. In order to demonstrate that let us
perform a simple Gaussian integral in (\ref{J}-\ref{SI0}) exactly
for the case of a free quantum particle with a mass $m$. Just for the
sake of convenience we will assume a continuous spectrum of the oscillators
\begin{equation}
\sum_{n}\frac{\pi C^2}{2\omega_n}[\delta (\omega_n-\omega )
-\delta (\omega_n+\omega )]= \eta \omega^{\beta} ,\;\;\;\; |\omega | <\omega_c,
\label{ohm}
\end{equation}
where $\omega_c$ defines the high frequency cutoff and $\beta$ is a
positive number. The exact results for the kernel $J$ are well known and
are presented elsewhere \cite{CL,HA,Gra,W}. Here we only focus on the most
important terms. One finds
\begin{eqnarray}
J \propto \exp\bigg[-mf_1(t)q_{-i}^2-mf_2(t)(q_{-f}-q_{-i})^2+...\bigg],
\label{Jint}
\end{eqnarray}
where $\gamma=\eta/m$ and  $q_{\pm i/f}$ are the initial/final values of $q_{\pm}$,
For the Ohmic bath ($\beta =1$) one obtains
\begin{eqnarray}
f_1(t)&=&\frac{\gamma}{2}\int\limits_0^t{\rm d}s \int\limits_0^t{\rm d}s'
\int\limits_{-\omega_c}^{\omega_c}\frac{{\rm d}\omega}{2\pi}
\omega \coth\frac{\omega}{2T}{\rm e}^{-{\rm i}\omega (s-s')}=
\gamma Tt+\gamma \ln\frac{1-{\rm e}^{-2\pi Tt}}{2\pi (T/\omega_c)}.
\label{f1}
\end{eqnarray}
In the long time limit the function $f_2$ coincides with the
equilibrium value of the average kinetic energy of the particle
$\langle E \rangle \equiv \langle m\dot q^2/2\rangle$:
\begin{equation}
f_2=\langle E\rangle=
\gamma\int\limits_0^{\omega_c}\frac{{\rm d}\omega}{2\pi}
\frac{\omega\coth\frac{\omega}{2T}}{\omega^2+\gamma^2}
\simeq
\frac{\gamma}{2\pi}\ln\frac{\omega_c}{\gamma}+
\frac{T}{\pi}\arctan\frac{T}{\gamma}.
\label{energy}
\end{equation}
We observe that the particle $q$
looses its coherence due to interaction with the
bath of oscillators. Indeed in the long time limit we have $f_1(t)\gg f_2$
and the kernel (\ref{Jint}) effectively reduces to
\begin{equation}
J \to \frac{1}{L}{\rm e}^{-m f_2q_{-f}^2}\delta(q_{-i}),
\label{limit}
\end{equation}
where $L$ is the system size. {\it Any} perturbation of the
density matrix will relax to the same equilibrium form
\begin{equation}
\rho(q_1,q_2)=(1/L){\rm e}^{-(q_1-q_2)^2/L_d^2}, \;\;\; L_d^2=1/mf_2
\label{dens}
\end{equation}
which is not sensitive to the initial phase. As a result of interaction
with the bath even at $T \to 0$ the off-diagonal elements of
the equilibrium density density matrix (\ref{dens}) decay on a typical length scale
$L_d \sim 1/\sqrt{\eta \ln (\omega_c/\gamma )}$ set by interaction.
The average value of the kinetic energy
of the particle $\langle E \rangle$ (\ref{energy}) is not zero even at $T=0$
irrespectively to its initial energy. At sufficiently low $T$ its value is
also determined by the interaction parameter $\gamma$. Another observation
is that the high frequency cutoff $\omega_c$ explicitely enters the expressions
(\ref{f1},\ref{energy}) which in the limit $T \to 0$ describe the
{\it true ground state properties}
of the system. The same results can be obtained within the imaginary time
technique, or just by an exact diagonalization of the initial Hamiltonian
of the system ``particle+oscillators'' \cite{HA}. From the latter work it is
particularly transparent that all the results depend on $\omega_c$ just
because {\it all} the oscillators (including the high frequency ones)
``take part'' in the diagonalization. By no means this implies excitation
of such oscillators. Rather one can say that in the presence of interaction
the noninteracting energy levels of the oscillators acquire
a finite width and they can exchange energy with a particle
in arbitrarily small portions. As a result of this exchange the particle
energy is distributed as
$$
w(E)=\int \frac{dp}{2\pi}\delta \left(E-\frac{p^2}{2m}\right) f(p)
\propto \exp (-E/2\langle E\rangle ),
$$
with the average value $\langle E\rangle$ given by eq. (\ref{energy}). Here
we defined
$$
f(p)= L \int dq_-\rho (q_-) e^{-ipq_-},
$$
where $\rho (q_-)$ is the equilibrium density matrix (\ref{dens}) and
$q_-=q_1-q_2$.

In order to avoid misunderstandings we would like to
emphasize that we (on purpose) work in basis of ``noninteracting''
eigenstates of the system. It is
obvious that the full wave function of the total system as well as
each of the eigenmodes obtained by an exact diagonalization always stay coherent.
However, since the behavior of the particle $q$ (and not that of the
eigenmodes) is of interest for us, the reduced density matrix $\rho (q_1,q_2)$
should be studied. The decay of the off-diagonal elements of $\rho$
on the length scale $\sim L_d$ just implies that the bath in
some conventional sense ``measures'' the particle position \cite{HA}.
In principle the off-diagonal elements of $\rho (q_1,q_2)$
(and thus the coherence of the particle $q$) can be suppressed
completely ($L_d$ tends to zero if one e.g. chooses $\omega_c \to \infty$),
while the eigenmodes of the total system obviously remain fully coherent.
Thus it is quite useless to discuss the presence or absence of
quantum coherence in the interacting many-body system without discussing
which quantity is actually calculated and/or measured in experiments.

Let us also note that the particular dependence of the results
on the high frequency cutoff $\omega_c$ is fully determined by the spectrum
of the bath (\ref{ohm}). E.g. for $\beta < 1$ for most of the physical
quantities of interest the cutoff does not enter at all, for $\beta =1$ this
dependence is logarithmic, while for $\beta >1$ one gets a power-law
dependence on $\omega_c$. In the latter case also the decay of
perturbations of the density matrix to its equilibrium value at $T=0$ is
faster than logarithmic. All these features have been extensively
studied \cite{CL,HA,Leg,Gra,W} and we will not go into more details here.

We conclude that the presented above Golden rule approach
fails to reproduce all the features obtained from the exact solution.
E.g. at $T=0$ the Golden rule approach yields $\langle E \rangle =\eta 0 =0$,
while the correct result is given by eq. (\ref{energy}), according to this
approach the initial state does not decay as a result of interaction,
while in reality it does (cf. (\ref{Jint}-\ref{f1})), according to
(\ref{Wif2}) no coherence can be lost at $T=0$, while actually it is
lost and even in equilibrium the off-diagonal elements of the density
matrix decay on a typical length scale $\sim L_d$
set by interaction. The Golden rule approach cannot correctly describe
both the ground state properties of an interacting system as well as their
low-lying excitations (cf. e.g. Ref. \onlinecite{HA} where such properties were
derived by means of an exact diagonalization).

The example with a free damped quantum particle is not unique, of course.
One can also consider e.g. a degenerate two-level system and observe that
interaction with the CL bath leads to the effective renormalization of
the transition amplitude $\Delta$ between the two levels \cite{Leg}:
$\Delta \to \Delta_r=\Delta (\Delta /\omega_c)^{\frac{\alpha}{1-\alpha}}$,
where $\alpha \propto \eta$. In the weak interaction limit one finds
$\Delta_r=\Delta (1- \alpha \ln (\omega_c/\Delta))$. Again this is the
property of the true ground state of an interacting system which
explicitely depends
on the high frequency cutoff $\omega_c$ even for very small $\alpha$. Again
it is impossible to obtain the above result for $\Delta_r$ within the
Golden rule approach which at $T=0$ gives $\Delta_r-\Delta = \alpha 0=0$.

Coming back to the problem of interacting electrons in a disordered metal
we observe exactly the same situation. The Golden-rule-type perturbation
theory \cite{AAG2} is not sufficient at low $T$ just because it cannot correctly
account for the ground state properties of an {\it interacting} system.
Proceeding perturbatively one starts from a noninteracting system and
imposes energy conservation. Obviously no dephasing can occur at $T=0$
because the electron energy $\langle E \rangle$ is zero in this case.
But this simply means that one keeps the energy of an interacting
system lower that it is in its true ground state.
In the presence of interaction this energy is {\it not} zero
$\langle E \rangle >0$ even at $T=0$. Although we do not
calculate this energy explicitely within our analysis it is obvious
that the scale for $\langle E \rangle$ is set by interaction and the
high frequency cutoff of the effective environment {\it will} enter. As in
the above example of the exactly solvable CL model this dependence
has nothing to do with excitation of high
frequency oscillators. This is just the property of the true ground
state of an interacting system.

Thus we see that our problem has essentially the same qualitative
features as the CL model. Of course, there exist also physical differences
between these problems. For instance, there seems to be no quantity in the CL
model which would be completely analogous to the decoherence time
$\tau_{\varphi}$ measured in the weak localization experiments. One can,
however, qualitatively compare the length $L_d$ in the CL model
with the decoherence length $L_{\varphi } \sim \sqrt{D\tau_{\varphi}}$.
Actually both quantities agree qualitatively if one establishes the
correspondence
between the interaction parameters and the bath spectra of these
problems. More importantly, in both problems quantum decoherence
appears as a result of energy exchange between the particle and the
effective environment. [We emphasize again that we are working in the
basis of noninteracting electrons which is obviously not the basis
of the eigenstates in the presence of interaction.]
In both problems the effect cannot be captured within
the Golden rule approach where this energy exchange is forbidden at $T=0$.

In order to illustrate this point again let us ``forbid'' this exchange
in our results by hand. Formally it implies that one should take the
limit $\omega \to 0$ in the expression for the dielectric susceptibility
of the effective environment in which case Im $1/\epsilon (\omega , k)$
vanishes. Obviously this procedure is not justified both physically and
mathematically, but we just use it for illustration. For the
CL model one can e.g. integrate over $s-s'$ in (\ref{f1}) in the infinite limits,
get $\delta (\omega)$ and as a result recover only the first term $\gamma Tt$
while the second will be missing. Then in the limit $T \to 0$ one would
observe no decay of the initial density matrix which would be incorrect.
Analogously at $T=0$ one would get $\langle E \rangle =0$ which would
be incorrect too.
The correct procedure is -- as it was done above -- to integrate over
frequencies keeping the time finite and only then to send it to infinity.

Analogously if one (just by hand) substitutes the time integral in eq. (71)
of Ref. \onlinecite{GZ2} by the delta function
$\delta (\omega)$ in order to provide the limit
Im $1/\epsilon (\omega , k) \to 0$ (no energy exchange) one obtains
$$
\frac1{\tau_{\varphi}} \sim T\frac{e^2}{\sigma_d}\int
\frac{d^d\bbox{k}}{(2\pi )^d\bbox{k}^2}
$$
The result is again zero at $T=0$, but again the $T$-independent term will
be missing as in the above case of the CL model. [Note, however, that
for large $T$ one can recover the correct result \cite{AAK} proceeding
in such a way. This is achieved if one cuts the integral at the
lower limit at $k \sim 1/L_{\varphi} \sim 1/\sqrt{D\tau_{\varphi}}$].
Although the above procedure is not rigorous it clearly illustrates the physical
difference between the results \cite{GZ2} and \cite{AAG2}.

\section{}

In order to complete the critique of our theory AAG presented a collection
of various experimental results and reported a strong disagreement with
our theory. In some 3d systems the disagreement was found to be
4 to 5 orders of magnitude. This must look as a strong argument
against our theory.

However, looking closer at the experimental data one observes that
in many systems chosen by AAG for comparison the mean free path is
several or even one (!) Angstrom. Such strongly disordered systems
are well beyond the applicability of our quasiclassical theory. Hence,
for such systems the comparison cannot be carried out at all. On top
of that, many systems have the granular structure. Our analysis cannot be
directly applied to such systems as well, for instance because the simple
formula for the noise spectral density
\begin{equation}
\frac{\langle|V_{\omega,k}|^2\rangle}{a^{3-d}}=
\frac{\omega\coth\frac{\omega}{2T}}{\sigma_d k^2},
\label{Drude}
\end{equation}
used in our theory works only for relatively small frequencies.
[Here $d$ is the effective
dimension and $a$ is a film thickness in 2d or a square root of a wire cross
section area in 1d.] In such systems the high
frequency cutoff $\omega_c$ in (\ref{Drude}) is determined not by the
elastic scattering rate but rather by the effective capacitance of metallic
grains. Examples will be analyzed in the first part of this Appendix.

In the second part of this Appendix we will discuss several experiments
in which the saturation of $\tau_{\varphi}$ was observed. All these
experiments are in a quantitative agreement with our theory. It appears
that no alternative explanation of these experiments exists at the moment.

It was always implied that our analysis is applicable for good metals
with high conductivity and relatively low elastic scattering rate. {\it
Then and only then} one can use a simple formula (\ref{Drude})
up to frequencies $\omega\sim 1/\tau_e$. In such complicated and strongly
disordered systems as granular percolating systems and metallic
glasses one typically has $1/\tau_e\sim 10^{15}\div 10^{16}$ Hz.
This corresponds to energies $\sim 10^5 K$ where the simple approximation
(\ref{Drude}) does not work due to various reasons.
In a general case the noise power spectrum is expressed by
the formula
\begin{equation}
\frac{\langle|V_{\omega,k}|^2\rangle}{a^{3-d}}=
{\rm Im}\left( \frac{-4\pi}{k^2\epsilon(\omega,k)}\right)
\coth\left(\frac{\omega}{2T}\right)
\label{noise}
\end{equation}
which should be used for the comparison with experiments.
Below we will demonstrate by how much the noise
can be reduced (and hence $\tau_{\varphi}$ increased) in many systems
discussed in \cite{AAG2} if one uses a realistic
model for $\epsilon (\omega ,k)$. For 1d and 2d systems some possible sources
of the noise reduction have been already analyzed in Sec. 4c of Ref.
\onlinecite{GZ2}.

Let us e.g. consider an experimental work \cite{ArGer} where the decoherence
time $\tau_{\varphi}$ was measured in 3d Cu granular
percolating systems. According
to \cite{AAG2} the inverse inelastic time in this system is
$1/\tau_e \sim 2\times 10^{15}$ Hz and the elastic mean free path (extracted
from the diffusion coefficient) is $l \sim 4\times 10^{-8}$ cm. It is clear
that our analysis cannot be applied to systems with such small values of $l$
because the quasiclassical approximation should not really work there. But for a
moment we ignore this fact and concentrate on the expression for $\epsilon
(\omega ,k)$. The authors \cite{AAG2} use our formula (81) from Ref.
\onlinecite{GZ2} obtained
from the Eq. (\ref{Drude}) and reported 5 orders of magnitude difference
with the measured value for $\tau_{\varphi}$. But the formula (\ref{Drude})
definitely cannot be applied for granular materials already at frequencies
much lower than $10^{15}$ Hz! The effect of grain capacitances at not
very low $\omega$ is crucially important and drastically changes the frequency
dependence of $\epsilon$. Using the standard model for a granular metal
one easily arrives at the conclusion that the high frequency cutoff $\omega_c$
in (\ref{Drude}) never exceeds $\omega_c \sim 1/RC$ and can be even smaller
if the grain self-capacitance dominates over the intergrain capacitance
(here $C$ is the typical grain capacitance and $R$ is the typical intergrain
resistance). In order to get a rough estimate for $\omega_c$ in this
case we make use of the typical grain size $\sim (1\div 3)\times 10^{-6}$ cm
and the resistivity $\rho \approx 7\times 10^{-5}$ $\Omega$ cm reported
by the authors \cite{ArGer}. A reasonable estimate for $R$ would be
$R \sim 100$ $\Omega$. Although the grain capacitance is difficult to
estimate from the grain size we can guess that the corresponding (renormalized)
charging energy $E_C^* \lesssim E_C=e^2/2C$ should not exceed $E_C^* \sim 0.1$ K
(most probably $E_C^*$ is even smaller because the effect of capacitance
renormalization should be quite strong in \cite{ArGer}).
This yields an upper limit estimate for the high frequency cutoff
$\omega_c \sim 10^{11} \div 10^{12}$ Hz (most probably even lower).
As a result the estimate for the decoherence
rate $1/\tau_{\varphi}$ extracted by AAG from our simple formula should be
reduced by a factor $\sim (\omega_c\tau_e)^{3/2} \sim 10^{-5} \div 10^{-6}$.
We see that already such a simple estimate allows to completely remove
5 orders of
magnitude disagreement reported by AAG for the experiment \cite{ArGer}.
An analogous conclusion can be reached in the case of other granular systems,
e.g. \cite{Gersh2}. By saying that we do not want to claim a good agreement
of our theory with these experiments. Rather we want to emphasize that
a simple approximation (\ref{Drude}) (and hence our simple formulas
derived on its basis) cannot be applied e.g. for granular systems at
frequencies exceeding $\omega_c \ll 1/\tau_e$. The whole theory should be
substantially modified in this case.

We believe that the same conclusion can be made for metallic
glasses. There the problem to construct a reasonable model
for the dielectric susceptibility appears to be even more complicated.
Such systems are usually strongly disordered and the elastic scattering
rate is very high, typically
$1/\tau_e\sim 10^{15}\div 10^{16}$ Hz. These values are
of the same order as the conductivity and the plasma frequency. They are
also comparable to the atomic interlevel spacing.
It is not quite clear to which extent the glasses can be considered as
homogeneous materials. In some cases granular structure was reported,
so that some properties of metallic glasses can be similar to those of
granular materials. These systems are clearly different from simple metals
in many respects and therefore the simple approximation (\ref{Drude})
can hardly be applied already at moderately high frequencies, certainly
well below $10^{16}$ Hz. As in the case of granular metals the effective
cutoff $\omega_c \sim 10^{12}\div 10^{13}$ Hz would allow to completely
remove the disagreement in all cases. But -- we repeat -- the systems
are very complicated and it is not clear whether our formulas can be
applied even at $\omega \lesssim \omega_c$.

Another very important condition used in our theory is
$p_Fl\gg 1$ or, equivalently, $\epsilon_F\tau_e\gg 1$, where $\epsilon_F$ is the
Fermi energy. Only provided this
condition is satisfied the quasiclassical diffusion picture can be applied.
Again in the experiments with 3d and 2d metallic glasses this condition is
{\it not} satisfied.  In all the experiments which cannot be described by our
formulas the parameter $\epsilon_F\tau_e$ is smaller than $5$. Thus the
simple quasiclassical approximation used in our paper \cite{GZ2} does not work in
these cases. An interesting illustration of this point can be extracted from
the experimental data \cite{Lin}. In this work the two 2DEG samples
with $\epsilon_F\tau_e \sim 3$ and one sample with
$\epsilon_F\tau_e \sim 45$ were studied.
For the strongly disordered samples the
measured $\tau_\varphi$ at low temperatures exceeds the value predicted by our
theory by the factor $\sim 200\div 300$, while for a weakly disordered sample
the agreement
between the theory and the experiment is very good
($\tau_\varphi^{\rm exp}(T_{\min}\sim 300{\rm mK})=1.5\times 10^{-10}$ sec,
$\tau_\varphi^{\rm theor}(T=0)=2\times 10^{-10}$ sec).

Yet another problem is the accuracy of independent measurements of the
parameters of such complicated systems as strongly disordered granular
materials and metallic glasses. Many authors estimate
the elastic scattering time $\tau_e$ using the value for the density
of states obtained within the free electron model. Komori et al. \cite{Komori}
extracted the value $\tau_e$ for disordered Cu films directly from
the measured magnetoresistance curves and arrived at much higher values
for $\tau_e$ than obtained by another method. If we use the estimate
\cite{Komori} we will immediately conclude that our theory works well
in the case of disoredered Cu films. Here we do not want to discuss
which way of evaluation of $\tau_e$ is better. Rather we would
like to emphasize that in many cases the accuracy of determination
of the system parameters is clearly insufficient for quantitative
comparison with theoretical predictions.

It is also interesting to note that in the experiments with 3d metallic glasses
\cite{Sahnoune,Bieri}
the saturation of $\tau_\varphi(T)$ was observed at
relatively high temperatures, $T\sim 1\div 4$ K depending on the experiment.
According to the authors, this saturation can be explained neither
by heating nor by the effect of magnetic impurities. If so, what could
be an alternative explanation for this effect?
If, following the
authors \cite{Sahnoune,Bieri}, one adopts a free electron model of
metallic glasses and
estimates the values of $\sigma$, $\tau_e$ and other parameters,
one immediately arrives at the conclusion that the simple Drude formula
(\ref{Drude}) is not applicable in this case and the noise should be
greatly reduced. If, furthermore, one assumes (again
following the authors \cite{Sahnoune,Bieri}) that in the metallic glasses
the effect of electron-phonon interaction is more important than
that of electron-electron interaction, then the
experimental data should be compared with the eq.(82) of Ref. \onlinecite{GZ2}.
The results are summarised in the Table I:

\begin{table}[h]
\begin{tabular}{cccccccc}
Sample & $1/\tau_e$, Hz & $p_Fl$ & $E_F\tau_e$ & $\sigma$, Hz & $c$, km/sec &
$1/\tau_\varphi^{\rm exp}(T=0)$, Hz & $1/\tau_\varphi^{\rm e-ph}(T=0)$, Hz \\
\hline
Cu$_{70}$Al$_{30}$\cite{Sahnoune} & $3\times10^{15}$ & 4 & 2 &
$2.9\times10^{15}$ & 6 & $1\times10^{10}$ & $6\times10^{10}$ \\
Cu$_{50}$Y$_{50}$\cite{Bieri} & $3.2\times10^{15}$ & 6 & 3 &
$5.7\times10^{15}$ & 2 & $4\times10^{9}$ & $1\times10^{10}$ \\
Y$_{80}$Si$_{20}$\cite{Bieri} & $1.1\times10^{16}$ & 2 & 1 &
$1.8\times10^{15}$ & 2 & $1\times10^{9}$ & $3\times10^{11}$ \\
\end{tabular}
\caption{}
\end{table}
Here $c$ is the sound velocity. Taking into account the uncertainity of the
experimental parameters as well as the accuracy of the theory, we can
conclude that the agreement between theory and experiment is good.

The electron-phonon formulas also well
describe the experiments in 2d Mg films \cite{White}. The results are given in
the Table II.

\begin{table}[h]
\begin{tabular}{cccccccc}
Sample & $1/\tau_e$, Hz & $p_Fl$ & $E_F\tau_e$ & $\sigma$, Hz & $c$, km/sec &
$1/\tau_{\varphi,\max}^{\rm exp}$, Hz &
$1/\tau_\varphi^{\rm e-ph}(T=0)$, Hz \\
\hline
Mg3 & $1.3\times10^{16}$ & 1.63 & 0.8 &
$1.6\times10^{15}$ & 6 & $9\times10^{9}$ & $6\times10^{10}$ \\
Mg4 & $6.1\times10^{15}$ & 3.55 & 1.8 &
$3.6\times10^{15}$ & 6 & $7\times10^{9}$ & $7\times10^{9}$ \\
Mg5 & $2.16\times10^{15}$ & 10 & 5 &
$1\times10^{16}$ & 6 & $1.5\times10^{8}$ & $5.8\times10^{8}$ \\
\end{tabular}
\caption{}
\end{table}
Again the agreement between the theory and the experiment is reasonable.
Thus in the above cases our theory provides a natural explanation for the
saturation of $\tau_{\varphi}$ at low temperatures. No alternative explanation
is known to us at the moment.

The saturation of the decoherence time extracted from the magnetoconductance
measurements was by now observed in very many experiments. Typically it
occurs already at relatively high temperatures of order 1 K. Although in
some cases it can be due to extrinsic factors and/or magnetic impurities
these reasons definitely cannot explain the saturation effect in all cases.
Just for an illustration let us consider three experiments from those quoted
in Ref. \onlinecite{AAG2}.

The decoherence time saturation was observed in \cite{Webb}.
In Ref. \onlinecite{AAG3} the authors argued that this saturation can
be due to the
effect of external noise. Let us not discuss this explanation but just
accept it for a moment. AAG concluded that the external noise may
effectively destroy the phase coherence without heating the sample
if the resistance of the latter is much smaller than the quantum
resistance 24 K$\Omega$. In the experiments \cite{Webb} the resistance
was smaller, so in that case the necessary condition was satisfied.

The decoherence time saturation was also observed in the experiments
\cite{Gersh}. In this case the resistance was at the M$\Omega$ level,
so that according to the authors \cite{AAG3} external noise cannot
be the reason for this saturation. At the same time the authors \cite{Gersh}
observed a rapid increase of the resistance which they interpreted as
a crossover to the strongly localized regime. As this crossover occurs close
to the temperature where $\tau_{\varphi}$ starts saturating the authors
\cite{Gersh} concluded that this saturation is not meaningful because
$\tau_{\varphi}$ makes little sense in the strongly localized regime.
Although this argument is essentially linked to the interpretation
of the resistance increase as a Thouless crossover, and although even
in this case we do not quite understand why the dephasing length
$L_{\varphi}$ extracted from the magnetoresistance measurements should
stay constant in the strongly localized regime being 3 times {\it smaller}
than the localization length, for a moment let us accept this argument
of the authors too.

The third experiment we are going to discuss is one by Pooke et al. \cite{Pooke}.
These authors also observed the decoherence time saturation in the
temperature range $T \lesssim 1$ K in three samples with resistances 120, 240
and 360 K$\Omega$. According to AAG \cite{AAG3} external noise cannot cause
this saturation (unless it produces overheating) since the resistance
of each of these 3 samples was much bigger than 24 K$\Omega$. Although the
authors \cite{Pooke} did not present the resistance curves one can hardly
expect the Thouless crossover to take place in their measurements in the
temperature range $0.1 \div 1$ K: e.g. for the 120 K$\Omega$ sample the
dephasing length $L_{\varphi}$ saturates at the level nearly 10 times
{\it smaller} than the localization length. We believe that under these conditions
it is already quite difficult to argue that in the experiments \cite{Pooke}
the ``saturated'' part of the curve
$\tau_{\varphi}(T)$ is meaningless. Since the effect of magnetic
impurities and heating are excluded in this case one should think about yet
another explanation of the effect specifically for the experiments \cite{Pooke}.
We are not aware of any proposal in this direction at the moment.

Now let us recall that the results of all 3 experiments \cite{Webb,Gersh,Pooke}
are in a quantitative agreement with our theory (see \cite{GZ1,GZ3}).
The parameters in these
experiments differ by several orders of magnitude, but for practically all
samples the agreement is within a numerical factor of order one. We believe
this agreement can hardly be interpreted as a simple coincidence. Rather we
can assume that this agreement strongly supports the validity of our calculation
which offers the same explanation for these three and many other experiments
and makes it unnecessary to search for a special explanation for each of
them separately. In some cases the agreement is very good,
for other experiments it is only qualitative, but this is rather a question
of finding a proper model for $\epsilon (\omega ,k)$ and/or the high
frequency cutoff
$\omega_c$ for the effective environment for this or that particular
system. Different
experiments are carried out on physically very different structures and it would
be naive to expect that there can exist a unique formula which alone could
quantitatively describe all existing weak localization experiments. But the
saturation effect at low temperatures appears to be a unique intrinsic
property of very many systems, and this effect is naturally explained by
our theory.

Finally, let us briefly discuss the experiments where the crossover to
the insulating behavior was observed. We believe that the experimental
results in 2d structures \cite{Imry,Liu,Hsu,Keuls} quoted by the
authors \cite{AAG2} by no means can be used as an argument against
our theory. In all these experiments the crossover to the insulating
state was observed for strongly disordered systems ($\epsilon_F\tau_e$ was
always of order one or only slightly larger). This is well beyond the
applicability range of our theory.

Furthermore, at least in
some cases (e.g. \cite{Liu,Hsu}) the films could have the granular
structure on the nanometer scale.
If this is the case the insulating crossover can be also explained in
terms of the Coulomb blockade of electrons on grains which does not
require any phase coherence of these electrons at all. It is well
known and was demonstrated experimentally (see e.g. \cite{CB}) that
this effect occurs in granular metals and arrays at temperatures
$T \lesssim E_C^*$, where $E_C^*$ is the effective (renormalized) charging
energy of the grain. If the resistance of a granular array or film (or,
equivalently, the intergrain resistance) is (roughly) above
the quantum resistance unit $R_q \approx 24$ K$\Omega$, the capacitance
renormalization is not important and the bare charging energy $E_C$ sets
the temperature scale for the metal-to-insulator crossover $T_{cr} \sim E_C$.
The estimate $E_C \gtrsim 10$ K is resonable for the grain size
in the nanometer range and is consistent with the experimental data. If
the intergrain
resistance is lower than $R_q$ the effective charging energy gets strongly
renormalized due to charge fluctuations (see e.g. \cite{SZ}). As a result
one gets $E_C^* \propto E_C \exp (-AGR_q)$, where $G$ is the film resistance
and $A$ is constant. For orderered square 2d granular arrays one has
$A=2$, and $A$ is definitely bigger in a disordered case. This might
also explain the
exponential dependence of $T_{cr}$ on $GR_q$ reported e.g. in
Ref. \onlinecite{Liu}. We can also quote recent theoretical
results by Nazarov \cite{Naz} who demonstrated that charging effects
may persist in disordered conductors even without grains and
tunnel barriers.  All these facts as well as some other experimental details
(like e.g. an obvious inconsistency of the experimental data \cite{Keuls} with
the standard scenario of the orthogonal-to-unitary transition in the
magnetic field) suggest that interactions can play quite an important role
in the above experiments. It is not our purpose to discuss this issue in
more details here. But in any case it is quite clear that the above
experimental data can hardly be interpreted as contradicting to our theory.

The same conclusion can be made concerning the experiments \cite{Gersh}
where the metal-to-insulator crossover in quasi-1d structures was reported
and interpreted as the evidence for the Thouless crossover to the
stronly localized regime at low $T$. It is clear, however, that the
``noninteracting'' scenario of this crossover cannot be applied in this case.
For instance, at the crossover temperature $T \sim 1$ K the
{\it measured} value of $T\tau_{\varphi}$ was found to be
$T\tau_{\varphi} \sim 0.3 \div 0.5$ depending on the sample and remained
of order one even
deep in the weak localization regime (e.g. at $T \sim 10$ K). As it was
already discussed above the dephasing length $L_{\varphi}$ was found to
saturate at the value 3 times smaller than the ``noninteracting'' localization
length. All that implies that interaction effects play a very important role
in the experiments \cite{Gersh}. This was also acknowledged by the authors
\cite{Gersh}. If so, the physical origin of the crossover can be debated.
It is not clear how the Thouless scenario should be modified in the
presence of strong interaction. There exist other mechanisms of localization
which are solely due to interaction. E.g. at $T=0$ a quantum particle can get
localized in a periodic potential if it interacts with an Ohmic bath of
oscillators \cite{Schmid}. In the subohmic case it is localized even in the
absence of any potential \cite{Gra}. Also the actual role of
Coulomb-blockade-type of effects in disordered conductors \cite{Naz}
should be understood better. All these effects do not involve the
phase coherence of electrons and therefore would be compatible with
the low temperature saturation of the decoherence length $L_{\varphi}$
predicted by our theory and found in experiments \cite{Gersh}.

Summarizing our discussion of the experiments we can conclude
that in most cases discussed in \cite{AAG2} the comparison with our
theory cannot be carried out at all because the corresponding
systems are very strongly disordered (the typical value of
the mean free path is in the range $l \sim 10^{-8}$ cm)
and therefore cannot be described by our quasiclassical theory. For
other experiments
the agreement is reasonable and can be improved further if a realistic
model for $\epsilon (\omega , k)$ is chosen. In many cases, especially for
quasi-1d systems, our theory is in a quantitative agreement with experiments.
Further experiments are needed to study the effect of low temperature
saturation of $\tau_{\varphi}$ in more details.


\begin{references}

\bibitem{AAG1} I. Aleiner, B.L. Altshuler, and M. Gershenzon,
cond-mat/9808078.
\bibitem{GZ1} D.S. Golubev and A.D. Zaikin, Phys. Rev. Lett. {\bf 81}, 1074
(1998).
\bibitem{GZ2} D.S. Golubev and A.D. Zaikin, cond-mat/9712203, Phys. Rev. B.
\bibitem{CL} A.O. Caldeira and A.J. Leggett, Physica {\bf A 121}, 587 (1983).
\bibitem{AAG2} I. Aleiner, B.L. Altshuler, and M. Gershenzon,
cond-mat/9808053.
\bibitem{HA} V. Hakim and V. Ambegaokar, Phys. Rev. B {\bf 32}, 423 (1985).
\bibitem{FV} R.P. Feynman and F.L. Vernon Jr., Ann. Phys. (NY) {\bf 24}, 118
(1963).
\bibitem{FH} R.P. Feynman and A.R. Hibbs, {\it Quantum Mechanics and Path
Integrals} (McGraw Hill, NY, 1965).
\bibitem{Leg} A.J. Leggett {\it et al.}, Rev. Mod. Phys. {\bf 59}, 1 (1986).
\bibitem{Gra} H. Grabert, P. Schramm, and G.L. Ingold, Phys. Rept. {\bf 168}, 115 (1988).
\bibitem{W} U. Weiss, {\it Quantum Dissipative Systems}, Series
in Modern Condensed Matter Physics, Vol. 2 (World Scientific, Singapore,
second edition, 1998).
\bibitem{CS} S. Chakravarty and A. Schmid, Phys. Rept. {\bf 140}, 193 (1986).
\bibitem{AAK} B.L. Altshuler, A.G. Aronov, and D.E. Khmelnitskii,
J. Phys. C {\bf 15}, 7367 (1982).
\bibitem{ArGer} A.G. Aronov, M.E. Gershenzon, and Yu. E. Zhuravlev, Sov. Phys.
JETP {\bf 60}, 554 (1984).
\bibitem{Gersh2} M.E. Gershenzon, V.N. Gubankov, and Yu.E. Zhuravlev, Sov. Phys.
JETP {\bf 58}, 167 (1983).
\bibitem{Sahnoune} A. Sahnoune, J.O. Strom-Olsen, and H.E. Fisher, Phys. Rev.
B {\bf 46}, 10035 (1992).
\bibitem{Bieri}  J.B. Bieri, A. Fert, G. Creuzet, and A. Schuhl, J. Phys.
F, Met. Phys. {\bf 16}, 2099 (1986).
\bibitem{Komori}  S. Komori, W. Kobayashi, and W. Sasaki, J. Phys.
Soc. Jap. {\bf 52}, 4306 (1983).
\bibitem{Lin} B.J.F. Lin, M.A. Paalanen, A.C. Gossard, and D.C. Tsui,
Phys. Rev. B {\bf 29}, 927 (1984).
\bibitem{White} A.E. White, R.C. Dynes, and J.P. Garno, Phys. Rev.
B {\bf 29}, 3694 (1984).
\bibitem{Webb} P. Mohanty, E.M.Q. Jariwala, and R.A. Webb, Phys. Rev. Lett.
{\bf 78}, 3366 (1997).
\bibitem{AAG3} I. Aleiner, B.L. Altshuler, and M. Gershenzon,
cond-mat/9803125.
\bibitem{Gersh} Yu.B. Khavin, M.E. Gershenzon, and A.L. Bogdanov, Phys. Rev.
Lett. {\bf 81}, 1066 (1998).
\bibitem{Pooke} D.M. Pooke, N. Paquin, M. Pepper, and A. Gundlach,
J. Phys. Cond. Mat. {\bf 1}, 3289 (1989).
\bibitem{GZ3} D.S. Golubev and A.D. Zaikin, cond-mat/9804156.
\bibitem{Imry} Z. Ovadyahu and Y. Imry, J. Phys. C {\bf 16}, L471 (1983).
\bibitem{Liu} Y. Liu, B. Nease, K.A. McGreen, and A.M. Goldman, Europhys.
Lett. {\bf 19} 409 (1992).
\bibitem{Hsu} S.-Y. Hsu and J.M. Valles, Phys. Rev. Lett. {\bf 74}, 2331 (1995).
\bibitem{Keuls} F.W. van Keuls, H. Mathur, H.W. Jiang, and A.J. Dahm, Phys.
Rev. B {\bf 56}, 13263 (1997).
\bibitem{CB} J.E. Mooij, B.J. van Wees, L.J. Geerligs, M. Peters, R. Fazio,
and G. Sch\"on, Phys. Rev. Lett. {\bf 65}, 645 (1990); T.S. Tighe,
M.T. Tuominen, J.M. Hergenrother, and M. Tinkham, Phys. Rev. B {\bf 47},
1145 (1993); P. Delsing, C. Chen, D.B. Haviland, and T. Claeson, Phys. Rev.
B {\bf 50}, 3959 (1994).
\bibitem{SZ} G. Sch\"on, and A.D. Zaikin, Phys. Rep. {\bf 198}, 237 (1990);
S.V. Panyukov and A.D. Zaikin, Phys. Rev. Lett., {\bf 67}, 3168 (1991).
\bibitem{Naz} Yu.V. Nazarov, cond-mat/9808340.
\bibitem{Schmid} A. Schmid, Phys. Rev. Lett. {\bf 51}, 1506 (1983).




\end{references}
\end{document}